# Applying cognitive diagnostic models to mechanics concept inventories

Vy Le,[1] Jayson M. Nissen,[2] Xiuxiu Tang,[3] Yuxiao Zhang,[3] Amirreza Mehrabi,[4]
Jason W. Morphew,[4] Hua Hua Chang,[3] and Ben Van Dusen[1]

[1]*School of Education, Iowa State University, Ames, Iowa 50011, USA*
[2]*Nissen Education and Research Design, Monterey, California 93940, USA*
[3]*College of Education, Purdue University, West Lafayette, Indianapolis 47907, USA*
[4]*School of Engineering Education, Purdue University, West Lafayette, Indianapolis 47907, USA*

In physics education research, instructors and researchers often use research-based assessments (RBAs) to assess students' skills and knowledge. In this paper, we support the development of a mechanics cognitive diagnostic to test and implement effective and equitable pedagogies for physics instruction. Adaptive assessments using cognitive diagnostic models provide significant advantages over fixed-length RBAs commonly used in physics education research. As part of a broader project to develop a cognitive diagnostic assessment for introductory mechanics within an evidence-centered design framework, we identified and tested the student models of four skills that cross content areas in introductory physics: apply vectors, conceptual relationships, algebra, and visualizations. We developed the student models in three steps. First, we based the model on learning objectives from instructors. Second, we coded the items on RBAs using the student models. Finally, we then tested and refined this coding using a common cognitive diagnostic model, the deterministic inputs, noisy "and" gate model. The data included 19 889 students who completed either the Force Concept Inventory, Force and Motion Conceptual Evaluation, or Energy and Momentum Conceptual Survey on the LASSO platform. The results indicated a good to adequate fit for the student models with high accuracies for classifying students with many of the skills. The items from these three RBAs do not cover all of the skills in enough detail, however, they will form a useful initial item bank for the development of the mechanics cognitive diagnostic.

## I. INTRODUCTION

Since the development of the Force Concept Inventory (FCI), research-based assessments (RBAs) have played an important role in shaping the landscape of physics education research [1,2]. RBAs have provided instructors and researchers with empirical evidence about how students learn and change throughout courses [1,3]. Researchers used data from RBAs to assess the impact of curricular and pedagogical innovations [1]. RBAs also play a central role in documenting inequities in physics courses before and after instruction [4,5]. In previous studies, researchers have primarily used the data from the RBAs as summative assessments to evaluate the effectiveness of a course [5–7].

Although some instructors use RBAs as formative assessments to inform their instructions, such as creating groups with diverse content knowledge [8,9], two shortcomings of existing RBAs hamper their use as formative assessments: (i) a lack of easily actionable information and (ii) a lack of timely information [3]. Examining overall RBA scores on student pretests can inform an instructor how well prepared a group of students is. Still, the overall RBA scores do not help instructors identify the specific skills students need to gain to be successful. Instructors and researchers also examine student gains in scores from the first (pretest) to the last (post-test) week of class. While this is a useful measure of the impact on instruction, it is an inherently retrospective activity that cannot inform instruction throughout a course.

To address the shortcomings of existing RBAs, we are developing the mechanics cognitive diagnostic (MCD). The MCD is a cognitive diagnostic (CD) computerized adaptive testing (CAT) assessment [10]. CD-CATs are adaptive assessments that can cover the specific contents and skills an instructor needs and wishes to assess them. CDs assess which skills a student has or has not mastered [11]. CATs can adapt to students' proficiency level and skill mastery profile, making assessment individualized and more efficient. These features allow an instructor to administer a CD-CAT as a formative assessment throughout a semester. The MCD will provide instructors with student-level and course-level assessments of student content knowledge and skill acquisition to help tailor instruction to students' needs.

To support the development of the MCD, we investigated the skills assessed by three RBAs commonly used in introductory college mechanics courses [1]. This research develops the models for the student skills and the evidence for assessing those skills as a component of the larger development of the MCD. The MCD will leverage this information to provide instructors with timely and actionable formative assessments.

## II. RESEARCH QUESTION

To support the development of the MCD to measure skills across introductory mechanics content areas, we developed and applied a model of four skills to three commonly used RBAs for introductory mechanics courses. To this end, we ask the following research question:

- What skills and content areas do three RBAs for introductory mechanics cover?

## III. DEFINITIONS

To support readers' interpretation of our research, Table I includes a selection of terms and their definitions.

## IV. LITERATURE REVIEW

Many physics education researchers and instructors use existing fixed-length RBAs. PhysPort [25] and the LASSO platform [26] provide lists and resources of these RBAs. Initially, instructors administered these RBAs with paper and pencil, but the administration is moving to online formats [27]. This move to online data collection has led to the development of CATs for introductory physics that have advantages over fixed-length tests. In this section, we discuss RBAs in introductory mechanics, options for administering RBAs online, CAT broadly, and the application of CAT to RBAs in physics.

### A. RBAs in introductory mechanics

PhysPort [28] provides an extensive list of RBAs for physics and other extensive pedagogical resources. PhysPort, however, does not administer assessments online. RBA developers and researchers have instead often relied on Qualtrics to administer the RBAs they develop or use online or the LASSO platform [26,27]. Administering RBAs online allows assessing students in class or outside of class to save class time, automatically analyze the collected data, and aggregate the data for research purposes [29].

PhysPort describes 117 RBAs [28] with 16 RBAs for introductory mechanics. Each RBA targets content areas and skills important for physics learning. The titles of each RBA often state the focus of the RBAs. For example, our study analyzed data from three RBAs because we had access to enough data for the analysis in this paper through the LASSO database. The Force Concept Inventory (FCI) [30] focuses on conceptual knowledge of forces and kinematics. The Force and Motion Conceptual Evaluation (FMCE) [31] provides similar coverage but has four energy questions. The Energy and Momentum Conceptual Survey (EMCS) [32] covers exactly what the name states. Other assessment names also portray skills or content areas of interest to physics education: the Test of Understanding Graphs in Kinematics, the Test of Understanding Vectors in Kinematics, and the Rotational Kinematics Inventory. These names imply that graphs and vectors play an important role in many physics courses and that many physics courses cover rotation. As discussed below, cognitive diagnostics allow for incorporating additional items to cover new topics throughout their lifetime.

TABLE I. Definitions of terms.

| Term—Definition |
| --- |
| Computerized adaptive testing (CAT)—Administered on computers, the test adaptively selects appropriate items for each person to match student proficiency [12–14]. |
| Proficiency—"…the student's general facility with answering the items correctly on the assessment under consideration" [15]. Higher proficiency increases the probability of answering assessment items correctly. Different fields use different terms for proficiency, such as skill, ability, latent trait, and omega. |
| Skills—A latent attribute that students need to master to answer items correctly and that cuts across content areas [13,16,17]. |
| Q-matrix—A Q-matrix, or "question matrix," is a binary matrix that maps the relationship between test items and the underlying skills they measure. Each row represents a test item, and each column represents a specific skill. An entry of 1 in the matrix indicates that a particular skill is required to answer the corresponding test item correctly, while a 0 indicates that the skill is not required. |
| Cognitive diagnostic (CD) assessment—An assessment method that evaluates students on specific skills to determine mastery. In contrast to traditional assessment methods that measure students on a single proficiency, CD provides diagnostic information on students' skill strengths and weaknesses to support personalized educational strategies [18,19]. |
| Classification accuracy—The agreement between observed and true skill classifications. In practice, this is calculated using the expected skill classifications rather than the true classifications, which is detailed in an example around Eqs. (4) and (6) in Ref. [20]. |
| Deterministic inputs, noisy "and" gate (DINA) model—A cognitive diagnostic model assuming that a student must master all the required skills to solve an item correctly. The absence of any required skills cannot be compensated by the mastery of others. This model operates within a binary framework, categorizing each skill as either mastered or not mastered [19,21–23]. |
| Evidence-centered design—A framework for developing educational assessments based on establishing logical, evidence-based arguments [24]. |

### B. Cognitive diagnostic—Computerized adaptive testing

Computerized adaptive testing (CAT) uses item response theory to establish a relationship between the student's proficiency levels and the probability of their success in answering test items [13]. CAT selects items based on student responses to the preceding items to estimate the student's proficiency and then aligns each item's difficulty with the individual's proficiency [13]. This continuous adaptation of item difficulty to student proficiency ensures that the test remains challenging and engaging for the students throughout its duration and provides a more precise estimation of the proficiency of students than paper-and-pencil assessment [12–14]. Compared to paper-and-pencil assessment methods, CAT requires fewer items to accurately measure students' proficiency meanwhile controlling the selected items concerning their content variety [33]. Chen et al. [34] show that CAT supports test security by drawing from a large item bank to control for item overexposure and how CAT can use pretest proficiency estimates for item selection and proficiency estimation to maximize test efficiency.

Combining cognitive diagnostic (CD) models and CAT improves the assessment process and categorizes students based on their mastery of distinct skills associated with each item. CD models aim to estimate how the students' cognitive proficiency relates to the specific skills or contents necessary to solve individual test items [13,35], with skill as a fundamental cognitive unit or proficiency that students need to acquire and master to answer certain items [16,17]. Deterministic inputs, noisy "and" gate (DINA) model emerges as a CD model that facilitates the assessment of skill mastery profiles and estimating item parameters [36]. The DINA model leverages a Q-matrix to test the relationships between items and the skills requisite to answer them [37], thereby providing a structured framework for monitoring the mastery levels of distinct proficiency [37]. The DINA model is applied for the evaluation of the mastery situation of students across various skills, including problem solving [38], computational thinking [17], and domain-specific knowledge [37].

### C. CAT in physics education

We are unaware of any CD assessments in physics. Researchers have, however, conducted studies on the effectiveness of CAT using an item response theory to evaluate students' proficiencies [12,39]. One such study by Istiyono et al. [40] utilized CAT to assess the physics problem-solving skills of senior high school students, revealing that most students' competencies fell within the medium-to-low categories. Morphew et al. [12] explored the use of CAT to evaluate physics proficiency and identify the areas where students needed to improve when preparing for course exams in an introductory physics course. Their studies showed that students who used the CAT improved their performance on subsequent exams. In another study, Yasuda et al. [41] also indicated CAT can reduce testing time by shorter test lengths while maintaining the accuracy of test measurement and administration. Yasuda et al. [39] examined item overexposure in FCI-CAT, employing pretest proficiency for item selection. This shortened test duration while maintaining accuracy and enhanced security by reducing item content memorization and sharing among students.

## V. THEORETICAL FRAMEWORK

We drew on evidence-centered design [24] to inform our development of the MCD. Evidence-centered design was first applied in the high-stakes contexts of the graduate record examinations [24,42] and has also been effectively utilized in physics education research for the development of RBAs [43,44]. We used three core premises in the evidence-centered design framework [24].

1. Assessment developers need content and context expertise to create high-quality items. In this analysis, we focused our analysis on three RBAs developed by physics education researchers—FCI, FMCE, and EMCS.
2. Assessment developers use evidence-based reasoning to evaluate students' comprehension and identify misunderstandings accurately. In this analysis, we developed a Q-matrix that identified which underlying skills were required to correctly answer each item (more details in Sec. VI B).
3. When creating assessments, developers must consider various factors such as resource availability, limitations, and usage conditions. For instance, the LASSO platform supports multiple-choice items and needs web-enabled devices, but it conserves class and instructor time.

Our work used the conceptual assessment framework provided by the evidence-centered design framework with its five models [24] (shown in Fig. 1) to guide assessment development. The models and their connections to our work are as follows:

1. *Student models* focus on identifying one or more variables directly relevant to the knowledge, skills, or proficiencies an instructor wishes to examine. In this project, a qualitative analysis (see Sec. VI B) indicated that four skills (i.e., apply vectors, conceptual relationships, algebra, and visualizations) and four content areas (i.e., kinematics, forces, energy, and momentum) would be optimum for our MCD.
2. *Evidence models* include evidence rules and measurement models to provide a guide to update information regarding a student's performance. The evidence rules govern how observable variables summarize a student's performance on individual test items. The measurement model transforms the student responses into the student skill profile. In this project, the evidence rules were binary, right or wrong scores, and the measurement model is the DINA model, which includes the Q-matrix.
3. *Task model* describes what students do to provide input to the evidence models. In this project, the *task model* was multiple-choice questions.
4. *Assembly model* describes how the three models above, including the student models, evidence models, and task models, work together to form the psychometric frame of the assessment. In the broader project, we developed a CD-CAT algorithm that integrated models 1–3 for the MCD.
5. *Delivery model* describes integrating all the models required for evaluation. We used the online LASSO platform [29,45] in this project.

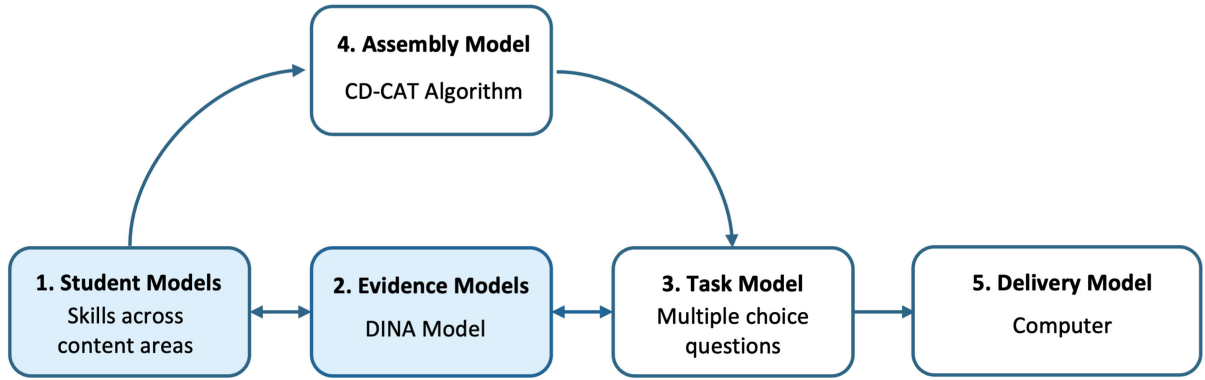

FIG. 1. An evidence-centered design framework for creating the mechanics cognitive diagnostic (MCD). This paper focuses on the student models and evidence models. The student models determine the skills and content areas that our assessment aims to measure. The evidence models apply the DINA model to the multiple-choice questions (task model) students answer to measure students' skills. Our CD-CAT algorithm will determine which items to ask students, who will take the assessment online through the LASSO platform.

In this paper, we focus on the student models and evidence models (models 1–2). These models are instrumental in aligning our analysis with the research question. By evaluating the student models, we gain insights into the range of competencies RBAs are designed to assess. Similarly, through the evidence models, we understand how these assessments capture and represent student understanding in various skills and content areas.

## VI. MATERIALS AND METHODS

To answer the research question, we employed a mixed methods approach using qualitative coding to identify the skills and content areas to measure for the student models. Subsequent quantitative analyses drove the testing of the evidence models and iterative improvements of the student models. We first used artifacts from courses to build the student models of skills that cut across the content of introductory mechanics courses. We then identified RBAs with sufficient data available through the LASSO platform and coded each item for the skills it assessed. Finally, we used an iterative process that applied the DINA model to build the evidence models and to improve our definitions of the skills and the coding of the skills on each item. In this iterative process, the DINA model suggested changes to the item skill codes initially made by content experts. The suggested changes were accepted or rejected by content experts. We then ran a final DINA model on our revised codes.

### A. RBAs data collection and cleaning

Our analysis examined student responses on three RBAs: the FCI (30 items, 12 932 students), FMCE (47 items, 5510 students), and EMCS (25 items, 1447 students). Our dataset came from the LASSO platform [26,29,45]. LASSO provided post-test data from 19 889 students across the three assessments. We removed assessments completed in less than 5 min and assessments with missing answers.

### B. Qualitative data analysis

We developed an initial list of skills and content areas covered in physics courses by coding learning objectives from courses using standards-based grading. We focused on standards-based grading because instructors explicitly list the learning objectives students should master during the course [46]. Initially, we coded a set of skills based on both the standards and the items on the RBAs; the skills included apply vectors, conceptual understanding, algebra, visualizations, and definitions. We discarded definitions as a skill because it represents a memorized response that the other skills covered in greater depth by asking students to apply or understand the concept. And, we are not aware of RBAs for introductory physics that ask definition questions. Table II lists the four skills and their definitions.

TABLE II. Definition of the skills in the FCI, FMCE, and EMCS assessments.

| Skills | Definition |
| --- | --- |
| Apply vectors | Item requires manipulating vectors in more than one dimension or has a change in sign for a 1D vector quantity. |
| Conceptual relationships | Item requires students to identify a relationship between variables and/or the situations in which those relationships apply. |
| Algebra | Item requires students to reorganize one or more equations. This goes beyond recognizing the standard forms of equations. |
| Visualizations | Item requires extracting information from or creating formal visualizations such as $xy$ plots, bar plots, or line graphs. |

TABLE III. Definition of the content areas in the FCI, FMCE, and EMCS assessments.

| Content areas | Definition |
| --- | --- |
| Kinematics | Items concerning the motion of objects without reference to the forces that cause the motion. |
| Forces | Free body diagram and Newtonian laws. |
| Energy | Conservation of energy, work, setup system, and the relationship between force and potential energy. |
| Momentum | Conservation of momentum and impulse. |

We initially coded content areas at a finer grain size to match the standards-based grading learning objectives, e.g., kinematics was split into four areas across two variables: 1D or 2D and constant velocity or constant acceleration. These content areas, however, were too fine grained to develop an assessment with a reasonable length for students to complete or a realistic size item bank. Therefore, we simplified the content codes: i.e., kinematics, forces, energy, and momentum for these three RBAs. Table III lists the four content areas covered by these three RBAs and their definitions.

Based on this initial set of codes we developed, we coded each item for its relevant skills and content areas. Our coding team included three researchers with backgrounds in physics and teaching physics. Each item was independently coded by at least two team members. The three coders then compared the coding for the items and reached a consensus on all items. This consensus coding of the three assessments provided one of the inputs into the DINA analysis.

### C. Quantitative data analysis
#### 1. DINA model

The Deterministic Input, Noisy "AND" gate (DINA) model is the foundational cognitive diagnostic model [21,22]. The DINA model is used to analyze responses to test items and determine the underlying skills that students possess [19]. A Q-matrix [47] (acting as a deterministic input) defines the relationship between test items and the required skills, which we defined in Table I. Each row of the Q-matrix corresponds to a test item, and each column corresponds to a skill. Q-matrix entries are binary, indicating whether a skill is needed for a specific item. The DINA model produces a skill profile for each student, represented as a binary vector, indicating whether they have mastered each skill. For example, a profile of [1, 0, 1, 0] means the student has mastered skills 1 and 3 but not skills 2 and 4. The DINA model assumes a student needs to have mastered all the required skills for a particular item to answer it correctly. If a student lacks even one required skill, the model assumes the student will answer the item incorrectly [23]. The model incorporates a probabilistic component (noisy "AND" gate) to account for real-world inconsistencies with two complementary parameters: slip ($s$) and guess ($g$). Slip is the probability that a student who has mastered all the required skills still answers the item incorrectly due to carelessness, distraction, or error. Guess is the probability that a student who has not mastered all the required skills answers the item correctly by guessing or other factors. Slip and guess add a stochastic element to help to account for the noise in real-testing scenarios, where students might guess or make unexpected errors. For each item, the probability that a student answers correctly is determined by whether they have the required skills and the slip and guess parameters. If the student has all required skills then $P(\text{correct}) = 1 - s$. If the student does not have all required skills then $P(\text{correct}) = g$. The model estimates each students skill profile based on their responses, the Q-matrix, and the slip and guess parameters for each item. We used the DINA model because the model fits indicated it was not necessary to use a more complex model like the generalized DINA model.

In this study, we used the DINA model to analyze students' response data for each of the three RBAs to refine our item codes further and calibrate each item's slip and guess parameters. The DINA model analyses also generated skill mastery profiles for each student, which were not the focus of the research question in this paper. These psychometric analyses were implemented using the G-DINA package [48] in the R programming environment. RMSEA2 and SRMSR were used to assess the degree of the model-data fit. RMSEA2 is the root mean square error approximation (RMSEA) based on the M2 statistic using the univariate and bivariate margins. RMSEA2 ranges from 0 to 1, and RMSEA2 < 0.06 indicates a good fit [49,50]. SRMSR, the standardized root mean squared residual, has acceptable values ranging between 0 and 0.8. Models with SRMSR < 0.05 can be viewed as a well-fitted model, and models with SRMSR < 0.08 are typically considered acceptable models [50–52]. Additionally, the skill-level classification accuracy, defined in Table I, informed the reliability and validity of the CD assessment. Classification accuracies range from 0 to 1, with values greater than or equal to 0.9 considered high [53,54] and values greater than 0.8 are acceptable [55].

The appropriateness of the Q-matrix plays an important role in CD assessments and affects the degree of model-data fit. Inappropriate specifications in the Q-matrix may lead to poor model fit and thus may produce incorrect skill diagnosis results for students. Therefore, we need a Q-matrix validation step in the study. The input Q-matrices for the DINA analysis for each RBA were constructed by content experts, as detailed in the prior section. In the Q-matrix validation step, detailed below, the DINA analysis further examined each Q-matrix to identify potential misspecifications in the Q-matrices.

### 2. Q-matrix validation

The analysis fitted the DINA model to students' post-assessment responses using the Q-matrix constructed by the three coders. The proportion of variance accounted for method [56] measured the relationships between the items and the skills specified in the provided Q-matrix. The analysis of the empirical response data suggested changes to the provided Q-matrix, which the three coders reviewed. The coders assessed the suggested modifications for how well they aligned with the definitions and revised the Q-matrix when the majority of the team agreed with the suggested changes. The refined Q-matrix was then used in subsequent CD modeling analyses.

Table IV presents a summary detailing the frequency of data-driven modifications suggested, adopted by the coders, and the rate of adoption for each of the three assessments under study. The FCI, for example, had 11 proposed changes of the 90 possible changes (30 items each with three possible skills), and the coders adopted 7 of these suggestions. For instance, conceptual relationships skill was initially not considered essential for item 7. However, empirical response data suggested that this skill was required to answer item 7 correctly. Postreview, the expert panel endorsed this modification; thereby, the value in the Q-matrix corresponding to the intersection of item 7 and conceptual relationships was changed from "0" to "1." Overall, only 8.5% of the codings (26 of 306) were identified for reexamination by this analysis. Of the 26 proposed changes, 13 were adopted across the 3 assessments, yielding an overall adoption rate of 50%. This iterative approach to informing the validity of the Q-matrix avoids overreliance on either expert opinion or empirical data, harmonizing both information sources to enhance the accuracy of the Q-matrix. Table VIII (see the Appendix) shows the final coding for each RBA item across the four content areas and four skills.

TABLE IV. Q-matrix modifications and adoption rates.

| | Total items | Possible changes | Suggested changes | Adopted changes | Adoption rate (%) | Change rate (%) |
|---|---|---|---|---|---|---|
| FCI | 30 | 90 | 11 | 7 | 64 | 7.8 |
| FMCE | 47 | 141 | 14 | 5 | 36 | 3.5 |
| EMCS | 25 | 75 | 1 | 1 | 100 | 4.0 |
| Overall | 102 | 306 | 26 | 13 | 50 | 4.2 |

## VII. FINDINGS

This section addresses the research question by detailing the skills and content areas measured by the three assessments, as detailed in Table VIII in the Appendix. First, we present which of the four skills the items on the three assessments measured and the number of skills the items measured. The specific models relating the items to the four skills are presented in the Appendix, see Tables IX–XI. Second, we show the content areas covered in the three assessments. Finally, we examine the skills across content areas. This structure highlights the various aspects of the items in these three assessments.

### A. Skills

FCI—The FCI assessed three skills (Fig. 2). Eighteen items assessed apply vector skill, 17 assessed conceptual relationships skill, 1 assessed visualizations skill, and 0 assessed algebra skill. The majority of items assessed a single skill. Twenty-four items (80%) assessed a single skill, 6 items (20%) assessed two skills, and 0 items assessed three skills (Table V).

FMCE—The FMCE assessed the same three skills as the FCI (Fig. 2). All 47 items assessed conceptual relationships skill, 19 items assessed the visualizations skill, 18 items assessed apply vectors skill, and 0 items assessed algebra skill. The majority of items assessed multiple skills. Thirteen items (28%) assessed a single skill, while 31

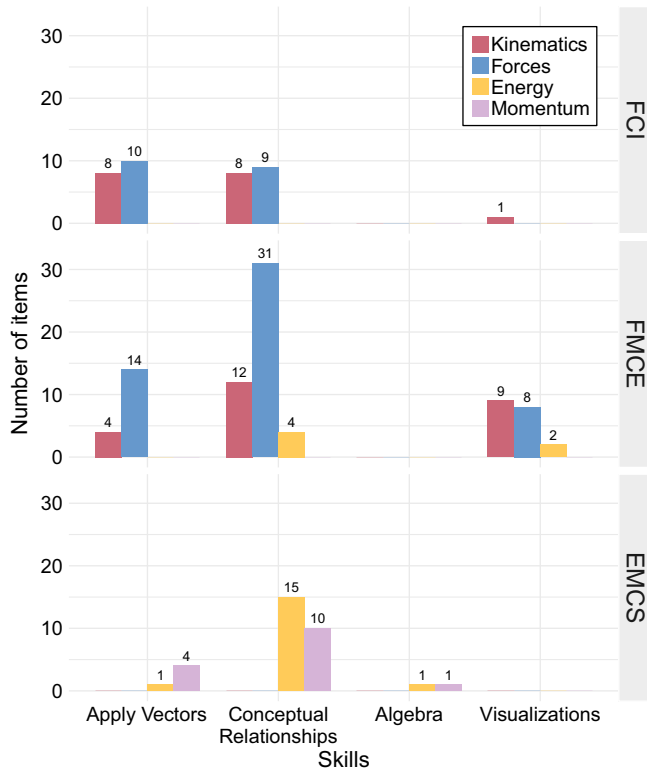

FIG. 2. The distribution of items across skills, content areas, and assessments. *Note that each item can assess multiple skills. Only 2 items, 3 and 13 of the EMCS assessed multiple content areas (i.e., energy and momentum) under the conceptual relationships skill.*

items (66%) assessed two skills, and 3 items (6%) assessed three skills (Table V).

EMCS—Similar to the FCI and FMCE, the EMCS assessed the apply vectors and conceptual relationships skills (Fig. 2). The EMCS differed in that it included 2 items that assessed the algebra skill. Of the 25 EMCS items, 23 assessed the conceptual relationships skill (with items 3 and 13 both coded for energy and momentum), 5 assessed the apply vectors skill, 2 assessed the algebra skill, and 0 assessed the visualizations skill. The EMCS was the only assessment with items assessing the algebra skill. The most of items assessed a single skill. Twenty items (80%) assessed a single skill, 5 items (20%) assessed two skills, and 0 items assessed three skills (Table V).

### 1. DINA model fit

The analysis fitted the DINA model with the refined Q-matrix to the response data. According to the established criteria [50,57], the model demonstrated satisfactory fit (RMSEA2 < 0.05, SRMSR < 0.07) for the FCI (RMSEA2 = 0.048, SRMSR = 0.062) and EMCS (RMSEA2 = 0.028, SRMSR = 0.041), whereas the fit for the FMCE was unsatisfactory (RMSEA2 = 0.090, SRMSR = 0.110). These outcomes suggest that the model adequately represents the underlying data structure for the FCI and EMCS but might not capture the latent structure of the FMCE well.

### 2. DINA model classification accuracy

Table VI presents the classification accuracy [20] for each skill across the three RBAs. As discussed in the skills section, not all skills were measured by each of the RBAs; 9 of 12 were possible. For those skills that were measured, 7 of the 9 classification accuracies were high (over 0.9). The classification accuracy of visualizations for the FCI (0.79) and algebra for the EMCS (0.63) was notably lower. The lower classification accuracy reflects the lack of items measuring these skills (Fig. 2).

### B. Content areas

FCI—The FCI assessed two content areas (Fig. 2 and Table VII). Eighteen items assessed forces, 12 assessed kinematics, and 0 assessed energy and momentum. All 30 items (100%) assessed a single content area.

FMCE—The FMCE assessed three content areas (Fig. 2 and Table VII). Thirty-one items assessed forces, 12

TABLE VI. Skill classification accuracy by assessment.

|  | Apply vectors | Conceptual relationships | Algebra | Visualizations |
|---|---|---|---|---|
| FCI | 0.97 | 0.96 | ⋯ | 0.79 |
| FMCE | 0.96 | 0.98 | ⋯ | 0.91 |
| EMCS | 0.94 | 0.95 | 0.63 | ⋯ |

TABLE V. The distribution of items across the number of skills they assess.

| | Number of skills | | |
|---|---|---|---|
| | 1 | 2 | 3 |
| FCI | 24 (80%) | 6 (20%) | 0 (0%) |
| FMCE | 13 (28%) | 31 (66%) | 3 (6%) |
| EMCS | 20 (80%) | 5 (20%) | 0 (0%) |
| Total | 57 (56%) | 42 (41%) | 3 (3%) |

TABLE VII. The number of items across content areas they assess. Only 2 items, 3 and 13 of the EMCS assessed multiple content areas (i.e., energy and momentum) under the conceptual relationships skill.

| | Total | Kinematics | Forces | Energy | Momentum |
|---|---|---|---|---|---|
| FCI | 30 | 12 | 18 | 0 | 0 |
| FMCE | 47 | 12 | 31 | 4 | 0 |
| EMCS | 25 | 0 | 0 | 15 | 12 |
| Overall | 102 | 24 | 49 | 19 | 12 |

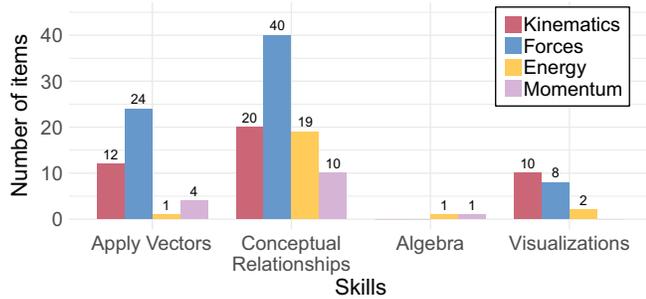

FIG. 3. The distribution of items from the three assessments (FCI, FMCE, and EMCS) across skills and content areas. Each assessment contains a different number of items, and some items assess multiple skills and content areas.

assessed kinematics, 4 assessed energy, and 0 assessed momentum. Similar to FCI, all 47 (100%) items assessed a single content area.

EMCS—The EMCS assessed two content areas (Fig. 2 and Table VII). Fifteen items assessed energy, 12 assessed momentum, and 0 assessed kinematics and forces. Unlike the FCI and the FMCE, 23 items assessed a single content area, and 2 items assessed two content areas (8%).

### C. Skills × content areas

The distribution of skills assessed was not consistent across content areas, see Fig. 3. This inconsistency follows from several aspects of the three RBAs. The FCI and FMCE did not measure the algebra skill. The EMCS did not measure the visualizations skill. Most items came from the FMCE and FCI, which focused more on forces than kinematics. Across the three RBAs, very few items measured the apply vectors skill for energy (1) and momentum (4), even though applying vectors is central to momentum. And, very few items measured the visualizations skills for energy (2) and momentum (0).

## VIII. DISCUSSION

This study supports the development of the MCD within the evidence-centered design framework by focusing on the student and evidence models (Fig. 1). For the student models, the three RBAs measured all four skills, though to different extents, across the four content areas. For the evidence models, the three RBAs assessed most of the skills with high classification accuracies. These results indicate that the combined items from the three RBAs will provide an adequate initial item bank for the further development of the MCD.

### A. Student models—The four skills

The three RBAs—FCI, FMCE, and EMCS—each included items that assessed three of the four skills across two to three content areas. The three RBAs all included a majority of items that assessed the conceptual relationships skill, which follows from their conceptual focus. In addition to measuring the conceptual relationships skill, all three RBAs also included sufficient items to assess the apply vectors skill with high classification accuracies. The FMCE included sufficient items to assess the visualizations skill. These results, in addition to other RBAs on visualizations and apply vectors specifically Zavala et al. [58], indicate that these three skills are common learning objectives of physics instruction.

The three RBAs did not include enough items assessing the algebra skill to inform how well that skill fits within our student models. This likely follows from these RBAs being conceptual assessments developed to refocus physics instruction from memorization and applying equations to a deeper understanding of the conceptual relationships linking the physical world. However, applying and manipulating equations was a common learning objective in the standards-based grading rubrics we used to develop our student models. Many instructors and students may want formative assessments on algebra skills to support their teaching and learning.

Most items in both the FCI and EMCS required mastery of a single skill, while most items in the FMCE needed multiple skills. Requiring multiple skills to answer an item correctly can have two effects. First, requiring mastery of more than one skill typically makes the items more difficult to answer. This is consistent with prior findings that the FMCE is more difficult than the FCI [6]. Second, multiskill items can provide different information than a single-skill item, and item banks should include a mix of single- and multiskill items to pick from to maximize the information generated by each item a student answers. Combining the three assessments into a single test bank provides a more even mix of single- and multiskill items than any of these three RBAs.

### B. Evidence models—Model fits and classification accuracies

The DINA model fit the FCI and EMCS well, but the fit for the FMCE was marginal. The length and difficulty of the FMCE may have driven this marginal fit. The large number of items assessing multiple skills may have also been a factor. *Post hoc* analyses to test these possibilities indicated that they were not major contributors to the marginal model fit of the FMCE. The additional analyses included the generalized DINA model, DINA models of the first and second half of the FMCE, and separate DINA models of the students from calculus- and algebra-based physics courses. The marginal fit likely follows from our *post hoc* application of our skill model to the FMCE. This model fits two assessments well, and one assessment marginally indicates that the student models of the skills are broadly applicable to physics learning, and items from these three assessments can form the initial item bank of a cognitive diagnostic.

The three RBAs had classification accuracy of above 0.9 for the apply vectors and conceptual relationships skills, as shown in Table VI. This makes sense for the FMCE and FCI, given that they each had at least 17 items for each of the apply vectors and conceptual relationships skills (Fig. 2). Although the EMCS only had 5 items measuring apply vectors, the classification accuracy was still 0.94. This finding indicates that a relatively small number of items can still accurately assess a skill. The number of items measuring algebra skills on the EMCS (2) and visualization skills on the FCI (1) was not sufficient to generate useful classification accuracies ($< 0.8$). Combining the three assessments into a single-item bank should provide sufficient coverage of apply vectors, conceptual relationships, and visualization skills, but it will not offer enough items to assess the algebra skill. Additionally, the combined item bank will require additional items to assess the visualization and apply vectors skills in the content areas of energy and momentum.

## IX. LIMITATIONS

The DINA analysis assumes students have mastered each skill assessed by an item to answer that item correctly. A less restrictive analysis, such as the generalized DINA, that assumes some questions can be answered by only mastering a subset of skills or by students who have only partially mastered skills may provide a better fit. The three RBAs constrained the skills that the analyses could test. This was an obvious issue for the algebra skill, which was only assessed by two items on one assessment. Physics instructors also likely value and teach other skills they would want to assess, such as the ability to decompose complex problems into smaller pieces to solve as assessed by the Mechanics Reasoning Inventory [59]). The analysis does not test the extent to which the items and assessments act differently across populations, e.g., gender, race, or type of physics course. Mixed evidence exists about the measurement invariance [60] and differential item functioning [61,62] of the FCI and FMCE. The combination of items from these three assessments administered through a cognitive diagnostic at a large scale will provide a dataset to identify and understand item differences and potential item biases between groups of students.

## X. CONCLUSIONS

Combining 102 items from three RBAs into a single item bank to create a CD-CAT provides a solid foundation for building the MCD. The limited number of items assessing the algebra skill and the apply vectors and visualizations skills for energy and momentum point to these as specific areas for improvement of the item bank. Delivering the MCD online, fortunately, has the advantage of allowing for the inclusion of new items under development to fill in gaps in the item bank. The combined item bank will also improve classification accuracy by having more items to draw on. However, the high classification accuracy (0.941) for the apply vectors skill on the EMCS indicates that even just 5 items can provide a high classification accuracy. This result indicates that shorter assessments may allow for high levels of classification accuracy for skills while also using fewer questions. We plan on ensuring sufficient classification accuracy with a minimum of ten items for each content area and skill combination. This will also provide enough items to estimate student proficiency when an instructor administers a single content area and skill combination as a weekly test. Future work will add content areas for mathematics and rotational mechanics.

Using LASSO as the delivery system for the MCD provides instructors with an adaptive tool to assess student's skills and knowledge across content areas or in specific content areas. In particular, using a cognitive diagnostic for the assembly model allows instructors to design formative assessments by choosing the skills and content areas to measure. Integrating guidelines and constraints on test lengths will help instructors design accurate assessments of those skills and content areas. The cognitive diagnostic also allows flexible timing; instructors can design pretests or post-tests that cover many skills and content areas or weekly tests focused on a few skills for one content area.

For researchers, the MCD will collect longitudinal data across skills and content areas. These data can inform the development of learning progressions or skills transfer across content areas, such as applying vectors in mathematical, kinematics, and momentum content areas. Developing more items that cover multiple content areas can inform how physics content interacts, which current RBAs do not assess. Because LASSO is free for instructors, the data will likely also represent a broader cross section of physics learners [63] than physics education research has historically included [64].

## ACKNOWLEDGMENTS

This research was made possible through the financial support provided by National Science Foundation Grant No. 2141847. We extend our appreciation to LASSO for their support in both collecting and sharing data for this research.

## APPENDIX

The Appendix includes the coding and refined Q-matrix tables (Table VIII–IX) for the three assessments used to conduct the DINA model analysis.

TABLE VIII. The skills and content areas for items from the FCI, FMCE, and EMCS. Note that "FCI_01" represents an abbreviation of the assessment name and the number of the item on the assessment.

| Content area | Apply vectors | Conceptual relationships | Algebra | Visualizations |
|---|---|---|---|---|
| Kinematics | FCI_07, FCI_08, FCI_09, FCI_12, FCI_14, FCI_21, FCI_22, FCI_23, FMCE_27, FMCE_28, FMCE_29, FMCE_41 | FCI_01, FCI_02, FCI_07, FCI_12, FCI_14, FCI_19, FCI_20, FCI_23, FMCE_22, FMCE_23, FMCE_24, FMCE_25, FMCE_26, FMCE_27, FMCE_28, FMCE_29, FMCE_40, FMCE_41, FMCE_42, FMCE_43 | | FCI_20, FMCE_22, FMCE_23, FMCE_24, FMCE_25, FMCE_26, FMCE_40, FMCE_41, FMCE_42, FMCE_43 |
| Forces | FCI_05, FCI_11, FCI_13, FCI_17, FCI_18, FCI_25, FCI_26, FCI_27, FCI_29, FCI_30, FMCE_01, FMCE_03, FMCE_04, FMCE_05, FMCE_06, FMCE_07, FMCE_08, FMCE_09, FMCE_10, FMCE_11, FMCE_12, FMCE_13, FMCE_20, FMCE_21 | FCI_03, FCI_04, FCI_06, FCI_10, FCI_15, FCI_16, FCI_24, FCI_25, FCI_28, FMCE_01, FMCE_02, FMCE_03, FMCE_04, FMCE_05, FMCE_06, FMCE_07, FMCE_08, FMCE_09, FMCE_10, FMCE_11, FMCE_12, FMCE_13, FMCE_14, FMCE_15, FMCE_16, FMCE_17, FMCE_18, FMCE_19, FMCE_20, FMCE_21, FMCE_30, FMCE_31, FMCE_32, FMCE_33, FMCE_34, FMCE_35, FMCE_36, FMCE_37, FMCE_38, FMCE_39 | | FMCE_14, FMCE_15, FMCE_16, FMCE_17, FMCE_18, FMCE_19, FMCE_20, FMCE_21 |
| Energy | EMCS_01 | EMCS_01, EMCS_02, EMCS_03, EMCS_04, EMCS_06, EMCS_08, EMCS_09, EMCS_12, EMCS_13, EMCS_15, EMCS_17, EMCS_20, EMCS_22, EMCS_24, EMCS_25, FMCE_44, FMCE_45, FMCE_46, FMCE_47 | EMCS_15 | FMCE_44, FMCE_45 |
| Momentum | EMCS_05, EMCS_11, EMCS_13, EMCS_23 | EMCS_03, EMCS_05, EMCS_07, EMCS_10, EMCS_13, EMCS_14, EMCS_16, EMCS_18, EMCS_19, EMCS_21 | EMCS_21 | |

TABLE IX. The table provides the refined Q-matrix for each FCI item, represented as binary coding, with * denoted adoption changes from the suggested Q-matrix of the DINA model.

| FCI item | Apply vectors | Conceptual relationships | Algebra | Visualizations |
|---|---|---|---|---|
| 1 | 0 | 1 | 0 | 0 |
| 2 | 0* | 1 | 0 | 0 |
| 3 | 0* | 1 | 0 | 0 |
| 4 | 0 | 1 | 0 | 0 |
| 5 | 1 | 0 | 0 | 0 |
| 6 | 0 | 1 | 0 | 0 |
| 7 | 1 | 1 | 0 | 0 |
| 8 | 1 | 0* | 0 | 0 |
| 9 | 1 | 0 | 0 | 0 |
| 10 | 0 | 1 | 0 | 0 |
| 11 | 1 | 0 | 0 | 0 |
| 12 | 1 | 1 | 0 | 0 |
| 13 | 1 | 0 | 0 | 0 |
| 14 | 1 | 1 | 0 | 0 |
| 15 | 0* | 1 | 0 | 0 |
| 16 | 0 | 1 | 0 | 0 |
| 17 | 1 | 0 | 0 | 0 |
| 18 | 1 | 0 | 0 | 0 |
| 19 | 0 | 1 | 0 | 0 |
| 20 | 0 | 1 | 0 | 1 |
| 21 | 1 | 0 | 0 | 0 |
| 22 | 1 | 0 | 0 | 0 |
| 23 | 1 | 1 | 0 | 0 |
| 24 | 0 | 1 | 0 | 0 |
| 25 | 1 | 1* | 0 | 0 |
| 26 | 1 | 0 | 0 | 0 |
| 27 | 1 | 0* | 0 | 0 |
| 28 | 0 | 1 | 0 | 0 |
| 29 | 1 | 0* | 0 | 0 |
| 30 | 1 | 0 | 0 | 0 |

(Table continued)

TABLE X. The table provides the refined Q-matrix for each FMCE item, represented as binary coding, with * denoted adoption changes from the suggested Q-matrix of the DINA model.

| FMCE item | Apply vectors | Conceptual relationships | Algebra | Visualizations |
|---|---|---|---|---|
| 1 | 1 | 1 | 0 | 0 |
| 2 | 0* | 1 | 0 | 0 |
| 3 | 1 | 1 | 0 | 0 |
| 4 | 1* | 1 | 0 | 0 |
| 5 | 1 | 1 | 0 | 0 |
| 6 | 1* | 1 | 0 | 0 |
| 7 | 1 | 1 | 0 | 0 |
| 8 | 1 | 1 | 0 | 0 |
| 9 | 1 | 1 | 0 | 0 |
| 10 | 1 | 1 | 0 | 0 |
| 11 | 1 | 1 | 0 | 0 |
| 12 | 1 | 1 | 0 | 0 |
| 13 | 1 | 1 | 0 | 0 |
| 14 | 0 | 1 | 0 | 1 |
| 15 | 0 | 1 | 0 | 1 |
| 16 | 0 | 1 | 0 | 1 |
| 17 | 0 | 1 | 0 | 1 |
| 18 | 0 | 1 | 0 | 1 |
| 19 | 0 | 1 | 0 | 1 |
| 20 | 1* | 1 | 0 | 1 |
| 21 | 1 | 1 | 0 | 1 |
| 22 | 0 | 1 | 0 | 1 |
| 23 | 0 | 1 | 0 | 1 |
| 24 | 0 | 1 | 0 | 1 |
| 25 | 0 | 1 | 0 | 1 |
| 26 | 0 | 1 | 0 | 1 |
| 27 | 1 | 1 | 0 | 0 |
| 28 | 1 | 1 | 0 | 0 |
| 29 | 1 | 1 | 0 | 0 |
| 30 | 0 | 1 | 0 | 0 |
| 31 | 0 | 1 | 0 | 0 |
| 32 | 0 | 1 | 0 | 0 |
| 33 | 0 | 1 | 0 | 0 |
| 34 | 0 | 1 | 0 | 0 |
| 35 | 0 | 1 | 0 | 0 |
| 36 | 0 | 1 | 0 | 0 |
| 37 | 0 | 1 | 0 | 0 |
| 38 | 0 | 1 | 0 | 0 |
| 39 | 0 | 1 | 0 | 0 |
| 40 | 0 | 1 | 0 | 1 |
| 41 | 1* | 1 | 0 | 1 |
| 42 | 0 | 1 | 0 | 1 |
| 43 | 0 | 1 | 0 | 1 |
| 44 | 0 | 1 | 0 | 1 |
| 45 | 0 | 1 | 0 | 1 |
| 46 | 0 | 1 | 0 | 0 |
| 47 | 0 | 1 | 0 | 0 |

TABLE XI. The table provides the refined Q-matrix for each EMCS item, represented as binary coding, with * denoting adoption changes from the suggested Q-matrix of the DINA model.

| EMCS item | Apply vectors | Conceptual relationships | Algebra | Visualizations |
|---|---|---|---|---|
| 1 | 1 | 1 | 0 | 0 |
| 2 | 0 | 1 | 0 | 0 |
| 3 | 0 | 1 | 0 | 0 |
| 4 | 0 | 1 | 0 | 0 |
| 5 | 1 | 1 | 0 | 0 |
| 6 | 0 | 1 | 0 | 0 |
| 7 | 0 | 1 | 0 | 0 |
| 8 | 0 | 1 | 0 | 0 |
| 9 | 0 | 1 | 0 | 0 |
| 10 | 0 | 1 | 0 | 0 |
| 11 | 1 | 0 | 0 | 0 |
| 12 | 0 | 1 | 0 | 0 |
| 13 | 1 | 1 | 0 | 0 |
| 14 | 0 | 1 | 0 | 0 |
| 15 | 0 | 1 | 1 | 0 |
| 16 | 0 | 1 | 0 | 0 |
| 17 | 0 | 1 | 0 | 0 |
| 18 | 0 | 1 | 0 | 0 |
| 19 | 0 | 1 | 0 | 0 |
| 20 | 0 | 1 | 0 | 0 |
| 21 | 0 | 1 | 1 | 0 |
| 22 | 0 | 1 | 0 | 0 |
| 23 | 1 | 0* | 0 | 0 |
| 24 | 0 | 1 | 0 | 0 |
| 25 | 0 | 1 | 0 | 0 |